# Disorder versus two transport lifetimes in a strongly correlated electron liquid

**Patrick B. Marshall[a]), Honggyu Kim, and Susanne Stemmer[b])**

Materials Department, University of California, Santa Barbara, CA 93106-5050, USA

[a)] Email: pmarshall@mrl.ucsb.edu

[b)] Email: stemmer@mrl.ucsb.edu

**Abstract**

We report on angle-dependent measurements of the sheet resistances and Hall coefficients of electron liquids in $SmTiO_3/SrTiO_3/SmTiO_3$ quantum well structures, which were grown by molecular beam epitaxy on (001) $DyScO_3$. We compare their transport properties with those of similar structures grown on LSAT [$(La_{0.3}Sr_{0.7})(Al_{0.65}Ta_{0.35})O_3$]. On $DyScO_3$, planar defects normal to the quantum wells lead to a strong in-plane anisotropy in the transport properties. This allows for quantifying the role of defects in transport. In particular, we investigate differences in the longitudinal and Hall scattering rates, which is a non-Fermi liquid phenomenon known as lifetime separation. The residuals in both the longitudinal resistance and Hall angle were found to depend on the relative orientations of the transport direction to the planar defects. The Hall angle exhibited a robust $T^2$ temperature dependence along all directions, whereas no simple power law could describe the temperature dependence of the longitudinal resistances. Remarkably, the degree of the carrier lifetime separation, as manifested in the distinctly different temperature dependences and diverging residuals near a critical quantum well thickness, was completely insensitive to disorder. The results allow for a clear distinction between disorder-induced contributions to the transport and intrinsic, non-Fermi liquid phenomena, which includes the lifetime separation.

**INTRODUCTION**

Strongly correlated materials exhibit transport properties that often strongly differ from normal metallic behavior [1-5]. Examples include poorly understood power laws in the temperature dependence of the resistance and a phenomenon known as lifetime separation, which was first discovered in the high-temperature superconducting cuprates [6,7]. It refers to the observation that the longitudinal resistance and the Hall angle (the inverse of the Hall mobility) exhibit different temperature dependencies and/or 0 K-limits, in stark contrast with the behavior of a simple, isotropic metal, which has a single relaxation time. In bulk materials, lifetime separation has been reported for cuprates [6,8,9], heavy Fermion systems [10,11], $V_2O_{3-x}$ [12], and ruthenates [13,14]. A direct consequence of the two distinct lifetimes is a temperature dependence of the Hall coefficient that is *not* due to a change in carrier density with temperature, i.e., a Fermi surface reconstruction; rather, it arises because it is the ratio of the two scattering rates. The fact that the same quasiparticle appears to exhibit two different scattering rates poses, however, significant theoretical challenges. Many different possible explanations have been proposed, including a two-dimensional Luttinger liquid (spin-charge separation), complex Fermi surfaces combined with strongly anisotropic scattering rates, as may occur near an antiferromagnetic transition, and quantum critical points [6,7,15-22].

Recently, lifetime separation has also been found in an interface-based correlated electron system [23]. These oxide heterostructures consist of electron liquids that are contained in thin $SrTiO_3$ quantum wells embedded in insulating rare earth titanates, i.e., $SmTiO_3$ [24]. The mobile sheet carrier density in the $SrTiO_3$ quantum well is fairly high (~ $7\times10^{14}$ cm$^{-2}$). It is fixed by the polar discontinuity at the $SrTiO_3/SmTiO_3$ interfaces [25] and thus independent of the quantum well thickness. The electron system in these quantum wells shows a number of correlation-induced phenomena, such as magnetism, pseudogaps, and transitions to an insulating state, as the three-dimensional carrier density is increased by decreasing the quantum well thickness [24,26-32].

The longitudinal resistance, $R_{xx}$, as a function of quantum well thickness could be described by [23]:

$$R_{XX} = R_0 + AT^n, \qquad (1)$$

where $R_0$ is the 0-K residual due to the disorder, $T$ is the temperature, $A$ is the scattering amplitude, and $n$ is the power-law exponent. For thick quantum wells, $n \sim 2$, similar to single interfaces [33]. Below a critical quantum well thickness, $t_{cr}$, as the system approaches an itinerant antiferromagnetic transition, $n < 2$ [26,29,34]. The Hall angle ($\theta_H$) is defined as:

$$H\cot\theta_H = H\frac{R_{XX}}{R_{XY}} = \frac{R_{XX}}{R_H} = H(C + \alpha T^2), \qquad (2)$$

where $H$ is the applied magnetic field, $R_{XY}$ is the Hall resistance, $R_H$ is the Hall coefficient, $C$ the residual (the inverse of the Hall mobility extrapolated to zero temperature), and $\alpha$ is the Hall scattering amplitude. Like $A$, $\alpha$ contains the effective mass, $m^*$, and it can be expressed as $\alpha T^2 = \frac{m^*}{e}\Gamma(T^2)$, where $\Gamma$ is a temperature-dependent scattering rate and $e$ the elementary charge. In contrast to $R_{xx}$, $\cot\theta_H$ remains $\sim T^2$, independent of the quantum well thickness. The residuals ($C$ and $R_0$) increased with decreasing quantum well thickness, consistent with increasing contribution from interface roughness scattering, as expected for thin quantum wells [35]. Furthermore, it was found that the residual $C$, but not $R_0$, contained an additional contribution near $t_{cr}$. The two observations, different temperature exponents and diverging residuals, define two separate transport scattering rates or lifetimes, as expressed in Eqs. (1) and (2), in close analogy to the other systems mentioned in the first paragraph above.

Disorder also plays an important role in the physics of strongly correlated systems. In some cases the presence of disorder can enhance the effects of electron correlations [36] and give rise to non-Fermi liquid behavior [37]. Introducing controlled amounts of disorder can also assist in elucidating the transport physics in correlated materials. For example, in the cuprates, it was shown that Zn impurities affect the 0 K-limit of the Hall angle scattering rate, implying that both lifetimes are associated with a real physical scattering process, i.e., not just an algebraic manipulation [6].

The goal of the study reported in this paper was to gain further understanding of the origins of lifetime separation by introducing defects into the $SrTiO_3$ quantum wells. We achieve this via planar growth defects normal to the quantum wells, which lead to strongly anisotropic transport properties. Using angle-dependent sheet resistance and Hall measurements, we determine the contributions of the defects and those of intrinsic scattering mechanisms to the transport properties and lifetime separation.

**RESULTS**

$SmTiO_3/SrTiO_3/SmTiO_3$ quantum well structures were grown on (001) $DyScO_3$ single crystals by hybrid molecular beam epitaxy (MBE) [38]. Like $SmTiO_3$, $DyScO_3$ belongs to the orthorhombic *Pbnm* space group. We note that our prior studies, the results of which were summarized above, were conducted on quantum wells grown on cubic LSAT [$(La_{0.3}Sr_{0.7})(Al_{0.65}Ta_{0.35})O_3$]. As will be discussed below, the substrates determine the crystallographic orientation of the $SmTiO_3$ barriers and impose epitaxial strains of different signs. The thicknesses of the bottom and top $SmTiO_3$ barriers were 10 nm. As in prior studies, we specify the SrTiO$_3$ quantum well thickness ($t_{QW}$) by the number of SrO layers it contains. Samples with SrTiO$_3$ quantum wells containing 2, 3, 4, 5, 6, and 10 SrO layers were investigated.

**Microstructure** High-angle annular dark-field (HAADF) scanning transmission electron microscopy (STEM) images along two mutually perpendicular directions of a sample containing a 6 SrO layer thick quantum well are shown in Fig. 1. The (001)$_O$ planes of the SmTiO$_3$ barrier are parallel to the substrate surface (the subscript refers to the orthorhombic unit cell), in contrast to previously investigated films on cubic LSAT substrates, for which SmTiO$_3$ grew in the (110)$_O$ orientation [39]. The (pseudo)-cubic lattice parameter of DyScO$_3$ is 3.95 Å [40], compared to the 3.87 Å of LSAT. The room temperature (pseudo-)cubic lattice parameter of SmTiO$_3$ is 3.91 Å [41], whereas that of SrTiO$_3$ is 3.905 Å [40]. Therefore, the sign of the epitaxial coherence strains for both the SrTiO$_3$ quantum well and the SmTiO$_3$ barrier differ on the two substrates (tensile on average for DyScO$_3$ and compressive for LSAT). The layers were coherently strained to the DyScO$_3$ substrate (see Supplementary Material). No extended defects are visible in the image taken along [010]$_O$, while along [100]$_O$ planar growth defects are detectable. These defects are normal to the growth plane and originate from the SmTiO$_3$/DyScO$_3$ interface. Their average spacing is ~ 20 nm. Similar defects and spacings were observed in the other samples (see Supplementary Material). A more detailed analysis of their atomic structure can be found in the Supplementary Material.

One may therefore expect anisotropy in the transport properties, in particular in the residual resistance, perpendicular and parallel to these defects. The A-site (Dy and Sm, respectively) displacements in the orthorhombic DyScO$_3$ and SmTiO$_3$ are clearly visible as a zigzag pattern normal to the surface along [100]$_O$. No such displacements are visible in the SrTiO$_3$ quantum well, indicating that there is no distortion to an orthorhombic structure that would be associated with oxygen octahedral tilts and A-site (Sr) displacements, at least not in the 6 SrO quantum well.

**Transport properties** The sheet resistances, $R_{XX}$, of Hall bar structures made from samples with 3 and 6 SrO layer thick quantum wells are shown in Figs. 2(a) and (b), respectively, along four in-plane directions, $[100]_O$, $[010]_O$, $[110]_O$, and $[1\bar{1}0]_O$. They exhibit clear anisotropies in both the residuals and temperature dependences. Using TEM, the direction with the highest resistance was identified as $[010]_O$. The higher resistance is consistent with increased scattering from planar defects, which are perpendicular to this direction. The lowest resistance direction, $[100]_O$, is parallel to the defects. The resistance along $[100]_O$ is similar to those of quantum wells on LSAT substrates [23], which show isotropic behavior and no planar defects in TEM. In contrast to the quantum wells on LSAT, the temperature dependence of $R_{XX}$ could *not* be described by a single power law, i.e. according to Eq. (1). This is particularly obvious in a derivative plot, $dR_{XX}/dT$ vs. $T$, shown in Fig. 2(c). A power law should result in a straight line in such a plot, which is clearly not the case.

The Hall angle is calculated from $R_{XX}$ and $R_H$, see Eq. (2), and one can extract two quantities: $HC$, the Hall angle residual (equivalent to the inverse of the Hall mobility extrapolated to zero temperature), and $H\alpha$, the temperature-dependent part of the scattering rate. Figure 3(a) shows the residual sheet resistance, $R_{XX}$ (2 K) and $HC$ as a function of angle with respect to the low resistance direction (defined as 0°) for the 6 SrO layer quantum well. Both $R_0$ and $HC$ vary with angle and are therefore, not unexpectedly, sensitive to the relative orientation between the current and the planar defects. The two quantities scale almost identically, i.e. disorder scattering due to the planar defects increases $R_0$ and $HC$ by the same factor. The dependence of $HC$ on disorder has previously been reported in the cuprates, where $HC$ was found to increase linearly with impurity dopant concentration [6,9]. Since it is also observed in the present study, this suggests that the linear relationship between *HC* and disorder is quite general and independent with respect to the nature of the defects. *Thus, as far as disorder scattering is concerned, there is only a single lifetime.*

In contrast, *Hα* is independent of disorder. This can be seen in Fig. 3(b), which shows $H \cot(\theta_H)$ of the 6 SrO quantum well as a function of $T^2$. The slope is determined by *Hα*. Here, $\cot(\theta_H) \sim T^2$, independent of the orientation and *Hα* shows almost no angular dependence, in contrast to *HC*, indicating that the anisotropy in transport is largely caused by the defects rather than by intrinsic anisotropy or strain (which would enter *Hα* through the effective mass). Furthermore, $\cot(\theta_H) \sim T^2$ applies also for the low resistance direction of the 3 SrO quantum well, which is also shown in Fig. 3(b) (the other directions for this sample were too resistive to obtain reliable measurements). The disorder-independence of *Hα* has also been noted previously in the cuprates [6,9]. *Thus, for the scattering mechanism that controls the temperature-dependent transport, there are two lifetimes, one that enters $R_{XX}$ and that does not have a simple power law T-dependence for these samples, and one that determines $\cot(\theta_H)$ and that is $\sim T^2$.*

Figure 4(a) shows *Hα* as a function of $t_{QW}$ alongside previous results [23] from quantum wells on LSAT. Similar trends are observed: a thickness-independent *Hα* followed by an abrupt increase below $t_{cr}$. The average *Hα* in thick wells is 50% higher on DyScO$_3$ ($1.5\times10^{-6}$ Vs/cm$^2$/K$^2$) compared to LSAT ($1\times10^{-6}$ Vs/cm$^2$/K$^2$). The critical thicknesses are 3 SrO layers on LSAT and 5 SrO layers on DyScO$_3$, while the upturn in the residual $R_{XX}$(2 K) occurs at one SrO layer on LSAT and three SrO layers on DyScO$_3$.

Figure 5 shows $(eR_H)^{-1}$ (in the absence of lifetime separation and for a single carrier type, it is equivalent to the sheet carrier concentration) of the 3 and 6 SrO layer thick quantum wells. Values along the high (blue) and low (orange) resistance directions are shown. At room temperature, $(eR_H)^{-1}$ corresponds to the expected $\sim 7 \times 10^{14}$ cm$^{-2}$. For the 6 SrO layer quantum well the temperature dependence is non-trivial [Fig. 5(b)]. This is a manifestation of carrier lifetime separation, because:

$$(eR_H)^{-1} = \frac{H}{e}\left(\frac{C+\alpha T^2}{R_0+AT^n}\right). \qquad (3)$$

A temperature-dependent $(eR_H)^{-1}$ arises if the low-temperature residual ratio, $C/R_0$, diverges and/or $n \neq 2$ in $R_{XX}$ and/or $R_{XX}$ shows a more complicated (non-power law) behavior as a function of $T$. The latter is the case here, as discussed above. The effect of a divergent $C/R_0$ is most evident at lower temperatures. It is not present for the 3 SrO layer quantum well, resulting in a nearly temperature-independent $(eR_H)^{-1}$. Remarkably, the carrier lifetime separation is the same along the high and low resistance directions, as the curves fall on top of each other. In other words, the contribution of the defects to the residuals $R_0$ and $C$ cancel each other [as was already evident in Fig. 3(a)], and $\alpha$ and $A$ are independent of disorder [as already seen in Fig. 3(b)].

Figure 6 shows the low temperature residuals $(eR_H)^{-1}$ (2 K), corresponding to $C/R_0$, of the quantum wells grown on LSAT (from ref. [23]) and DyScO$_3$ as a function of $t_{QW}$. In both cases, a divergence of $C/R_0$ ($C$ is the diverging quantity, see ref. [23]) is seen as a maximum in $(eR_H)^{-1}$ (2 K). It occurs at a slightly higher $t_{cr}$ (~ 6 SrO layers) for the quantum wells on DyScO$_3$.

**DISCUSSION**

We first compare the transport of the quantum wells in this study with our previous data on LSAT substrates, reported in ref. [23]. We then discuss the implication of the results on the phenomenon of lifetime separation.

**Effects of quantum well barrier orientation and epitaxial strain.** The quantum wells on the two substrates, $DyScO_3$ and LSAT, differ in the orientation of the $SmTiO_3$ barriers (they are 90º rotated relative to each other) and the epitaxial strain. Comparing $H\alpha$ in the quantum wells grown on $DyScO_3$ vs. those on LSAT (Fig. 4), we see that the qualitative behavior is very similar. $H\alpha$ is, however, larger in the quantum wells on $DyScO_3$. A likely origin is a larger in-plane effective mass (which enters $H\alpha$) associated with the tensile strain. The increase in $H\alpha$, observed in both types of samples at low $t_{QW}$, occurs at a larger value of $t_{QW}$ on $DyScO_3$. For very thin quantum wells, the Ti-O-Ti bond angle distortions were previously found to appear also in the $SrTiO_3$ quantum well [30,39,42,43], which may increase the effective mass and $H\alpha$ below $t_{cr}$. Theoretical and experimental studies also show that the magnetic order of the barrier couples into the itinerant electron system for sufficiently thin quantum wells [26,28,29,31,34], which may also increase the effective mass [44].

It remains an open question why $R_{XX}$ does not follow a well-defined power law, as was the case for the quantum wells on LSAT and single interfaces [23,33]. It is possible that these quantum wells are closer to a localization transition, influenced by the defects and the aforementioned mass enhancement. The influence of the defects on $R_{XX}(T)$ seems to be particularly obvious in the high resistance directions of the 3 SrO layer quantum well. Whatever the origin, it does not influence the temperature dependence of $\cot(\theta_H)$, which robustly remains $\sim T^2$.

**Implications for lifetime separation.** Disorder scattering appears in the residuals of the Hall angle and the longitudinal resistance. However, disorder has no influence on the lifetime separation. It is present in diverging residuals near $t_{QW} \sim 6$ SrO layers, causing the peak in the data in Fig. 6. Furthermore, the temperature dependence of the two scattering rates remains different. The most striking observation is that $\cot(\theta_H)$ remains robustly $\sim T^2$, even as the longitudinal resistance does not follow a simple power law.

Several of the models in the literature focus on the case where $R_{XX} \sim T$ and $\cot(\theta_H) \sim T^2$, as observed in the cuprates; the results presented suggest a more universal behavior of $\cot(\theta_H) \sim T^2$, independent of any particular $T$-dependence of $R_{XX}$. Prior results in ref. [23] for very similar structures, but grown on LSAT, showed that the power law exponent $n$ in $R_{XX}(T)$ correlated with different kinds of (incipient) magnetic order, whereas here, the departure of $R_{XX}(T)$ from a simple power law is not well understood but could reflect proximity to localization. In both cases, however, $\cot(\theta_H) \sim T^2$. It is thus tempting to attribute the $T$-dependence of $\cot(\theta_H)$ to that of a fundamental underlying intrinsic scattering mechanism that is active in a wide range of correlated materials (see below), whereas the specific behavior of $R_{XX}(T)$ is much more material specific. *A main conclusion is therefore that the ubiquitousness and robustness of the $T^2$ dependence of $\cot(\theta_H)$ should be an important consideration in developing models that describe correlated systems.* A similar argument has recently been made for the cuprates [45], but the present data indicates that the case can be made for a much wider class of materials.

A $T^2$-dependence of the resistance is often (but not always, see, e.g., ref. [46]) ascribed to electron-electron scattering in a Fermi liquid (in conjunction with an appropriate momentum relaxation mechanism). We have recently argued that a carrier density independent scattering rate observed in these quantum wells is not consistent with a Fermi liquid [47]. The carrier density independent scattering rate is also apparent in the data presented here. Figure 4(b) shows that $H\alpha$ is independent of $t_{QW}$ for $t_{QW} > t_{cr}$, even though narrowing the quantum well increases the *three-dimensional* carrier density. In a Fermi liquid, the carrier density enters the scattering rate through a (model dependent) interaction energy (see, e.g., Eq. (6) in ref. [1]). For this reason we believe that $\cot(\theta_H) \sim T^2$ should not be taken as evidence that there is an underlying Fermi liquid (the contrary appears to be true). We note that a carrier-density-independent scattering rate is found in *many other correlated materials* in the regimes where transport shows a scattering rate that is $\sim T^2$ [19,48-51]. In addition to electron-electron scattering, a $T^2$-dependent scattering rate can originate from magnetic fluctuations but also from orbiton excitations [13,52]. Given that the $T^2$-scattering rate is also observed in *non-degenerately* doped bulk $SrTiO_3$ [33,47,53-55], a non-magnetic origin seems more likely.

The results provide at least two intriguing hints as to the microscopic parameters that determine the Hall scattering rate. We have previously associated the diverging $C/R_0$ with a quantum critical point, i.e., zero-point fluctuations that appear in the residual $C$, but not in $R_0$, see ref. [23]. Like the previous studies, the present data show that the diverging $C/R_0$ is *not* due to disorder and also that it is likely not associated with a sudden change in electronic structure. For example, mass enhancement and effects of disorder are all seen in various transport parameters, as discussed above, but none show a peak at 6 SrO layers, in contrast to $C/R_0$. This seems to indicate that the Hall scattering rate, and thus perhaps also its temperature dependence, is fundamentally originating from strong electronic correlations. Secondly, as mentioned above, results may indicate that $\cot(\theta_H)$ is much less sensitive to carrier localization than $R_{XX}$. While two carrier models (i.e. the presence of two types of carriers, localized and non-localized [56]) cannot explain the data (see discussion in ref. [23]) it seems that the same carrier can appear more or less localized, depending on the transport coefficient interrogated.

**CONCLUSIONS**

To summarize, lifetime separation has two origins, namely a robust $T^2$-dependence of $\cot(\theta_H)$ and a divergence in the 0-K Hall residual, which, in the system studied here, occurs near a critical quantum well thickness. We showed that $\cot(\theta_H) \sim T^2$ persists independent of the degree of disorder and the specific temperature dependence of the longitudinal resistance. We speculated that the robust $T^2$ behavior of $\cot(\theta_H)$ reflects a universal, underlying scattering rate that is fundamentally due to electron correlations but *does not* indicate electron-electron scattering in a Fermi liquid. Modification of various materials parameters, such as epitaxial strain, introduction of anisotropic defects, and proximity to different types of magnetism, affect the longitudinal resistance but not the temperature dependence of the Hall angle scattering rate. The results, along with other experimental studies presented in the literature, emphasize the importance of accounting for these particular characteristics in the lifetime separation in theoretical models of non-Fermi liquid behavior in strongly correlated systems. They seem to describe an increasingly wide range of materials and thus indicate highly universal behavior.

**METHODS**

Details of the MBE growth approach were reported elsewhere [38]. Electrical contacts to the as-grown samples were made with electron-beam evaporated Ti/Au contacts in van der Pauw geometry, which have previously been demonstrated to remain Ohmic down to 2 K, the lowest temperature used in this study. After initial van der Pauw sheet resistance ($R_{XX}$) and Hall measurements were completed, standard photolithographic and etching processes were used to pattern Hall bar ("centipede") devices on the 3 and 6-SrO layer thick quantum well samples in a geometry which allowed transport measurements to be made along four crystallographic directions: [100], [010], [110], and [1$\bar{1}$0]. The transport measurements were performed from 2 to 300 K in a Quantum Design Physical Property Measurement System. The Hall resistance was measured with magnetic field sweeps from -0.6 to 0.6 T. For STEM studies, cross-sectional specimens were prepared using an FEI Helios dual beam focused ion beam microscope and imaged using a FEI Titan S/TEM operating at 300 kV.

**ACKNOWLEDGEMENTS**

The authors thank Evgeny Mikheev for many discussions and the development of the photolithography mask used during processing and Ryota Shimizu for development of the etching procedure used to treat the substrates. We also thank Jim Allen for many helpful discussions. P.B.M. acknowledges support through an NSF Graduate Fellowship. We acknowledge support from the U.S. Army Research Office (W911NF-14-1-0379). Acquisition of the oxide MBE system used in this study was made possible through an NSF MRI grant (Award No. DMR 1126455). The work made also use of the Central facilities supported by the MRSEC Program of the U.S. National Science Foundation under Award No. DMR 1121053.

## Author contributions

P.B.M. performed the film growth, processed the devices and carried out the electrical measurements and data analysis. H. K. performed the STEM analysis. P.B.M. and S.S. wrote the manuscript.

## Additional information

Supplementary information accompanies this paper at http://www.nature.com/ scientificreports

**Competing financial interests:** The authors declare no competing financial interests.

**Figure Captions**

**Figure 1:** HAADF-STEM images of a sample with a 6 SrO thick quantum well along (a) $[010]_O$ (Left) and (b) $[100]_O$ (Right). The SrTiO$_3$ quantum appears darker in these images, due to the lower atomic number of Sr compared to Sm. Planar defects [see red arrows in (b)] are visible along $[100]_O$ but not $[010]_O$ and have a relatively uniform spacing of approximately 20 nm. The schematics in (c-d) show the corresponding cross-sectional views of the DyScO$_3$. Note the pronounced zig-zag pattern of the A-site cations in (d), visible in the STEM image along $[100]_O$ for both DyScO$_3$ and SmTiO$_3$, which has the same orthorhombic structure, but are not seen in the SrTiO$_3$ quantum well.

**Figure 2:** $R_{XX}$ along four crystallographic directions measured as a function of temperature using a centipede structure for (a) a 3 SrO layer thick quantum well and (b) a quantum well containing 6 SrO layers. (c) Derivative plot, $dR_{XX}/dT$ vs. $T$, of the data shown in (b). The dashed line in (c) indicates the expected behavior for a $T^2$ temperature dependence. The labels indicate the transport directions.

**Figure 3:** (a) Low-temperature residuals of the Hall angle, $HC$ (blue), and sheet resistance $R_{XX}$(2 K) (orange) for the 6 SrO layer thick quantum well measured with the centipede device. The data is plotted as a function of the angle with respect to the low resistance direction, which is defined as 0°. (b) Hall angles of the 6 SrO layer thick quantum well along four directions plotted as a function of $T^2$, and for the low resistance direction of the 3 SrO layer thick quantum well (dashed line).

**Figure 4:** (a) Hall angle scattering amplitude $H\alpha$ plotted as a function of $t_{QW}$. For comparison, data from ref. [23] for quantum wells grown on LSAT substrates is also shown. (b) Low temperature residual sheet resistance, $R_{XX}$ (2 K) as a function of $t_{QW}$.

**Figure 5:** $(eR_H)^{-1}$ as a function of $T$ of the highest resistance (blue) and lowest resistance (orange) directions for (a) the 3 SrO layer (b) the 6 SrO layer thick quantum well.

**Figure 6:** Low temperature residual of $(eR_H)^{-1}$ at 2 K as a function of $t_{QW}$. For comparison, data from ref. [23] for quantum wells grown on LSAT substrates is also shown.

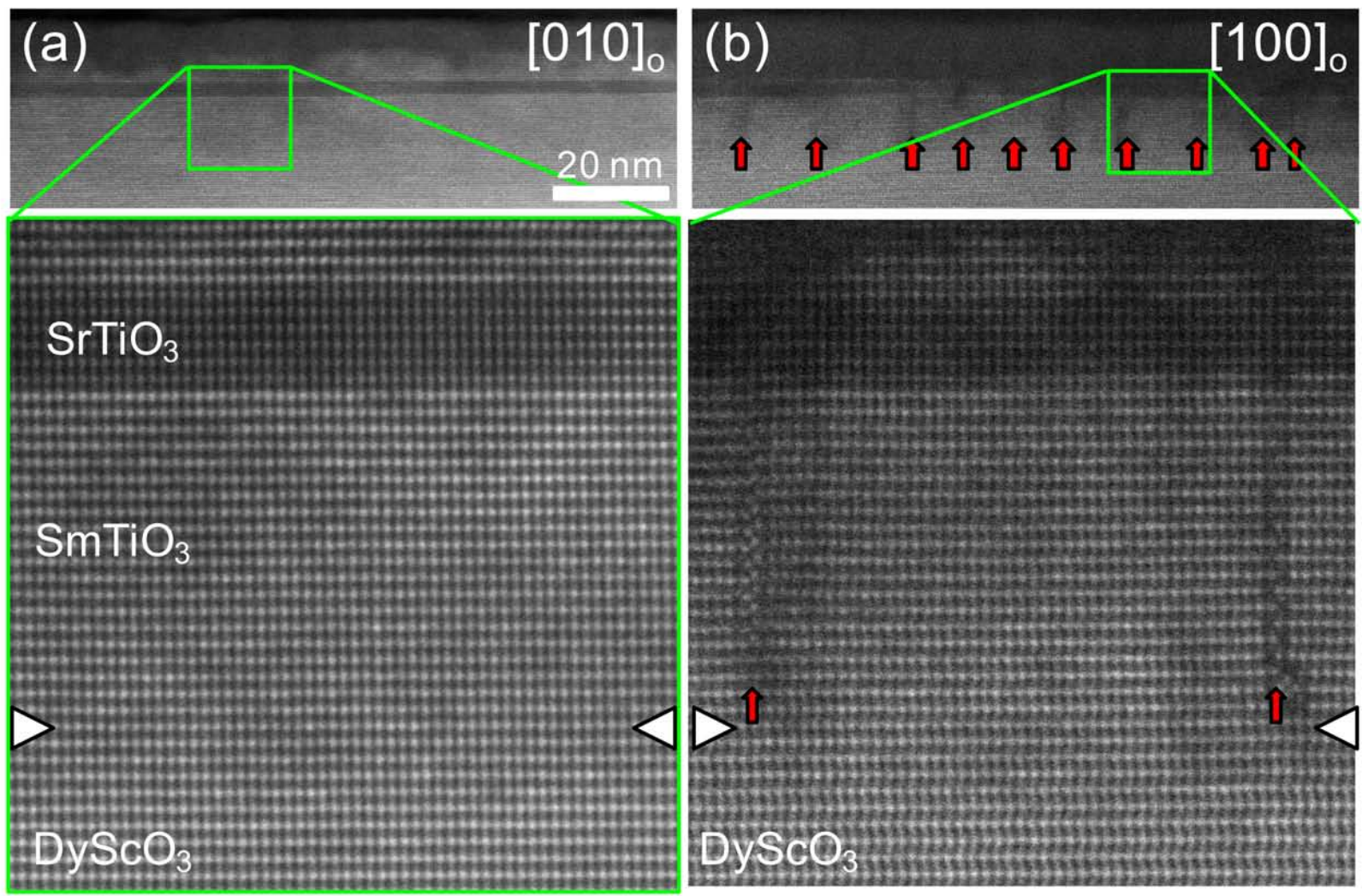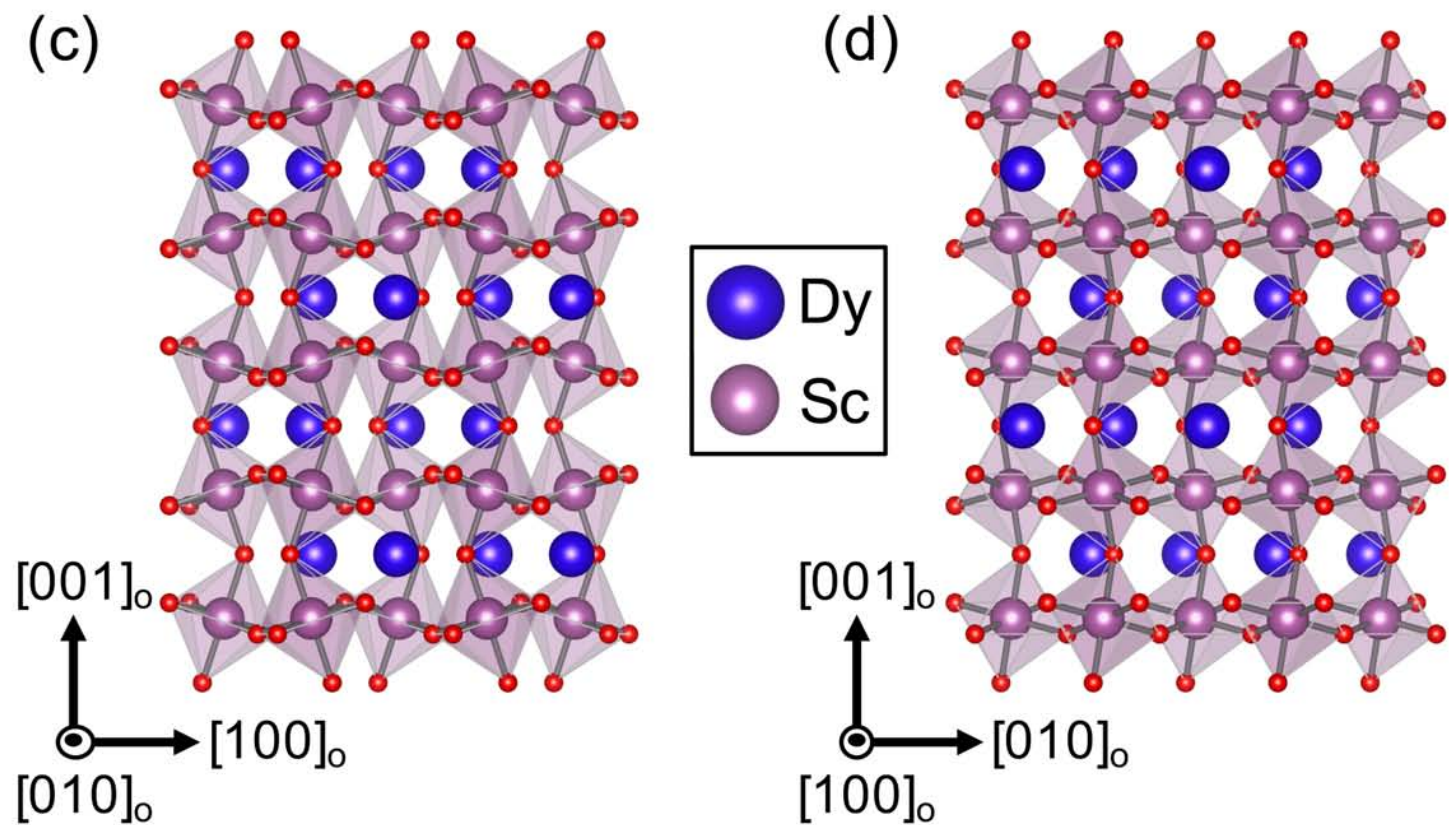

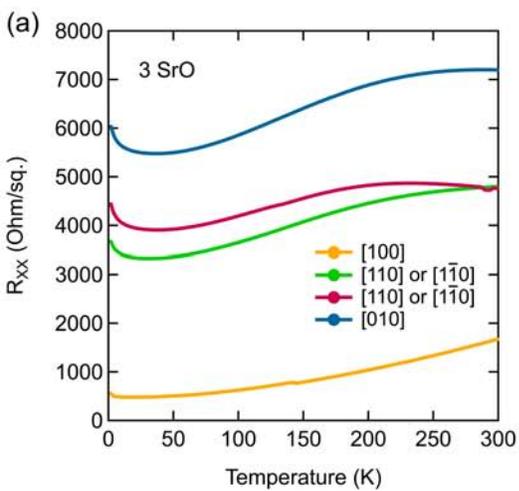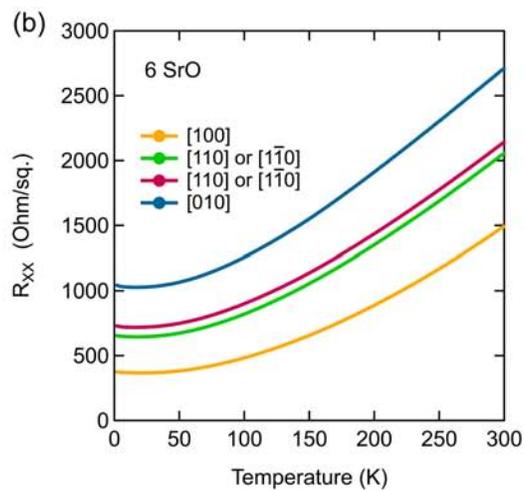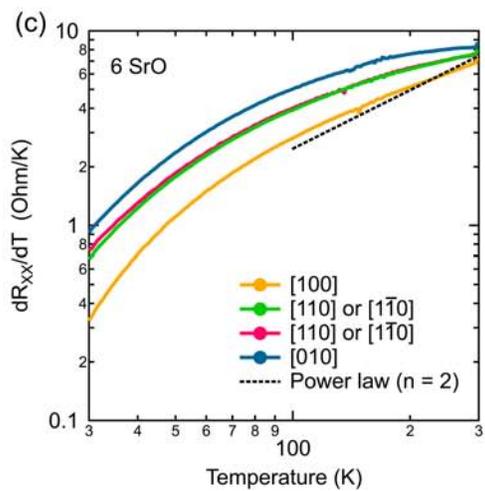

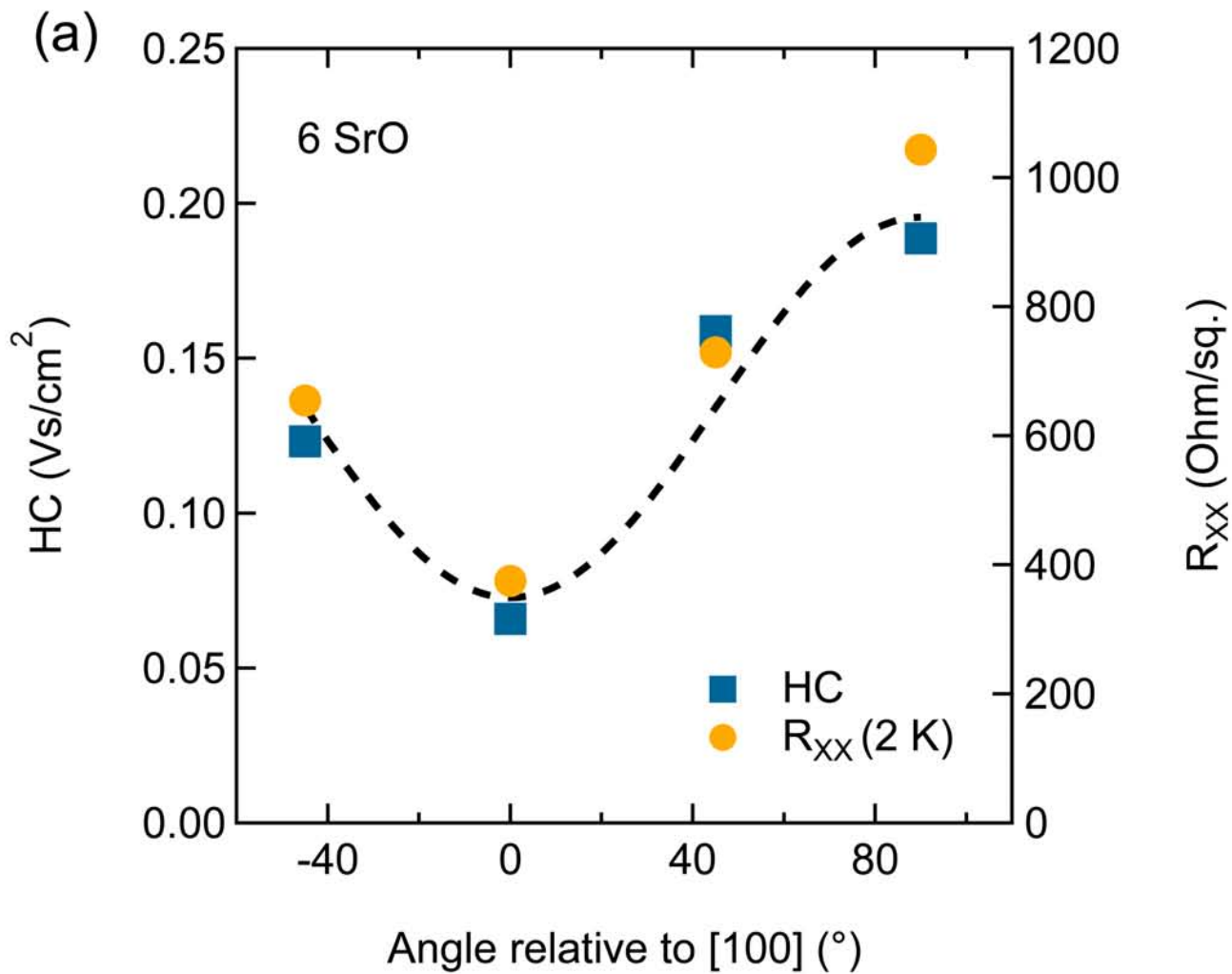
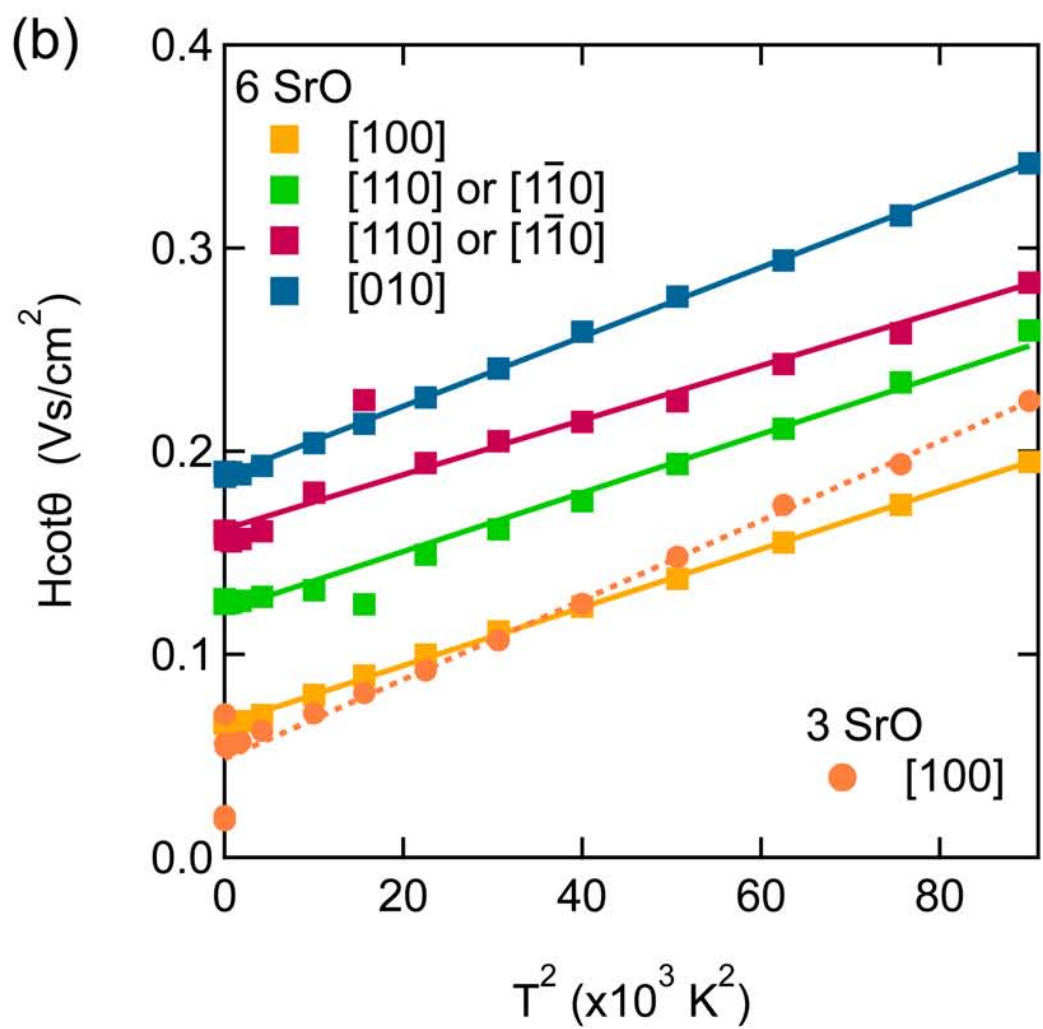

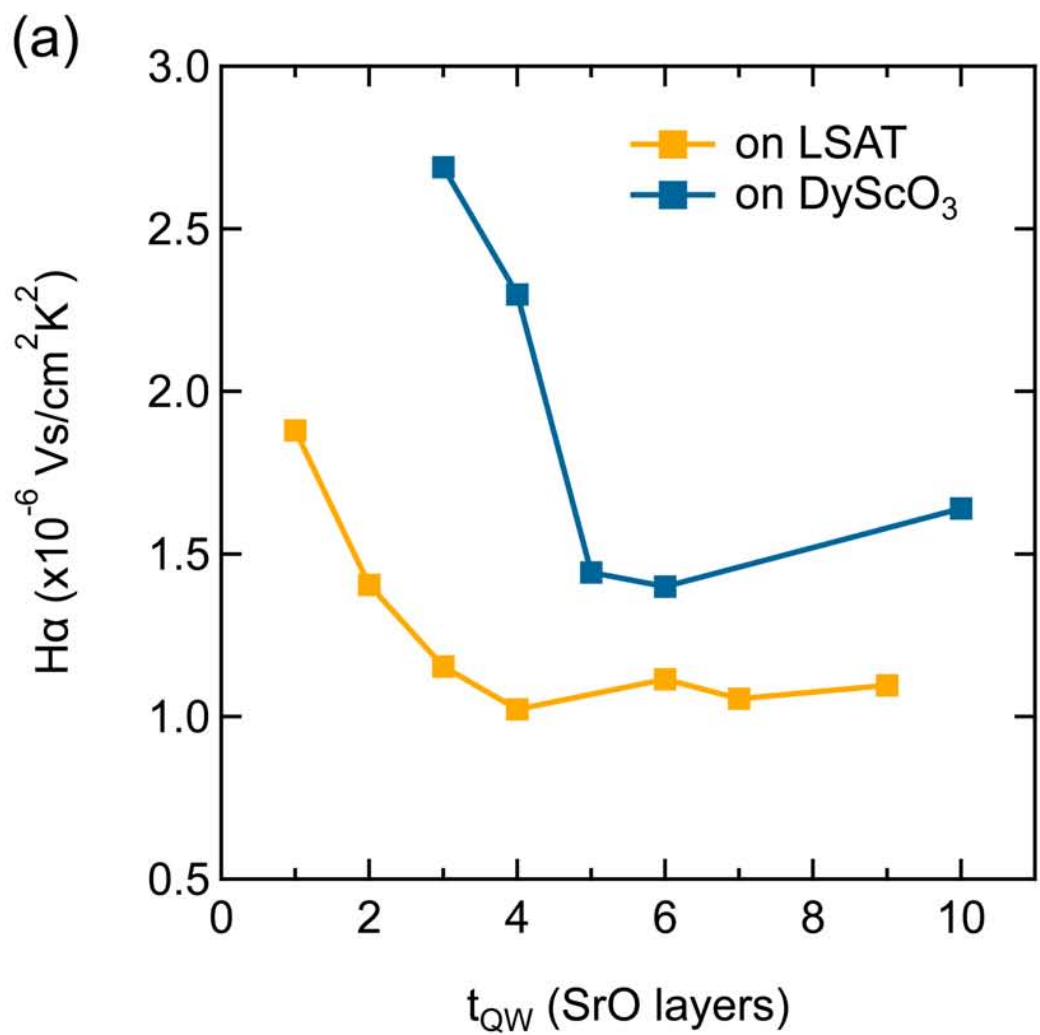

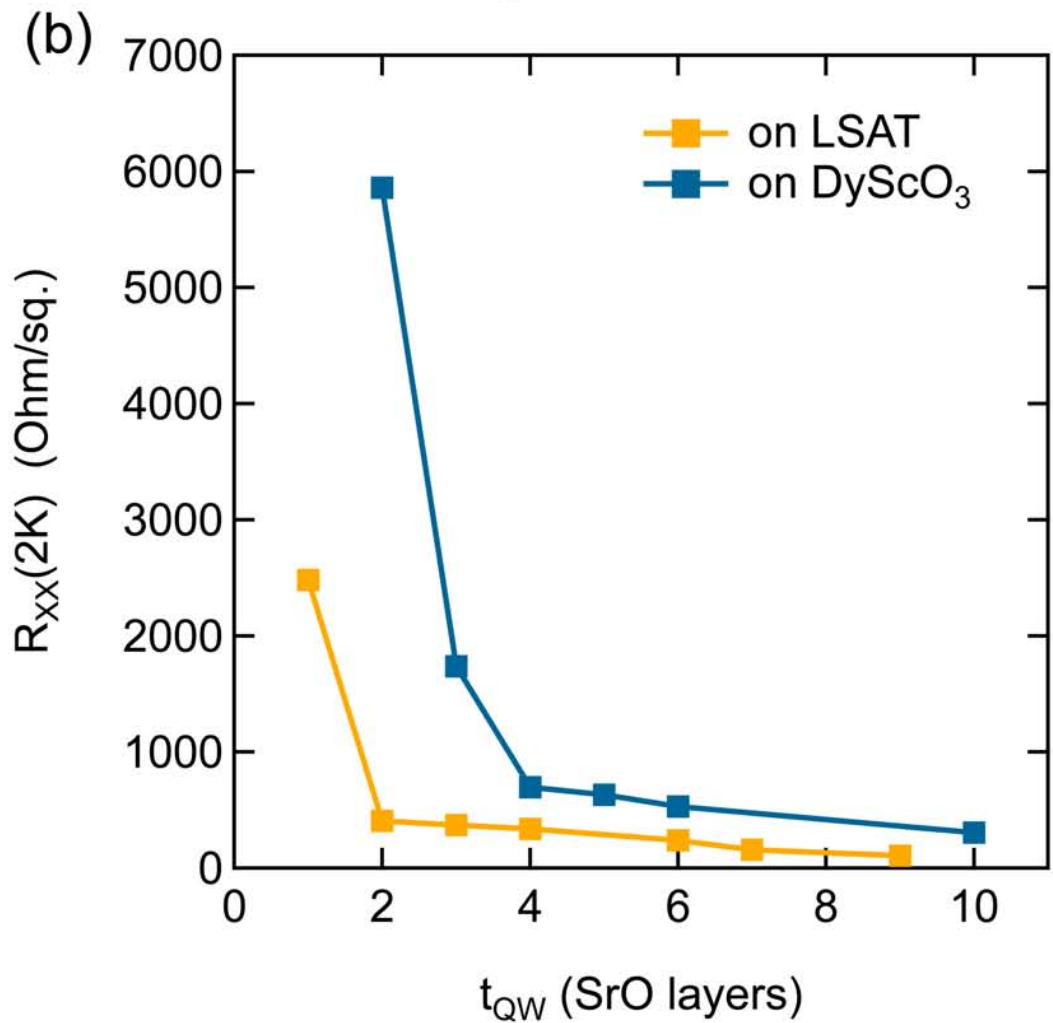

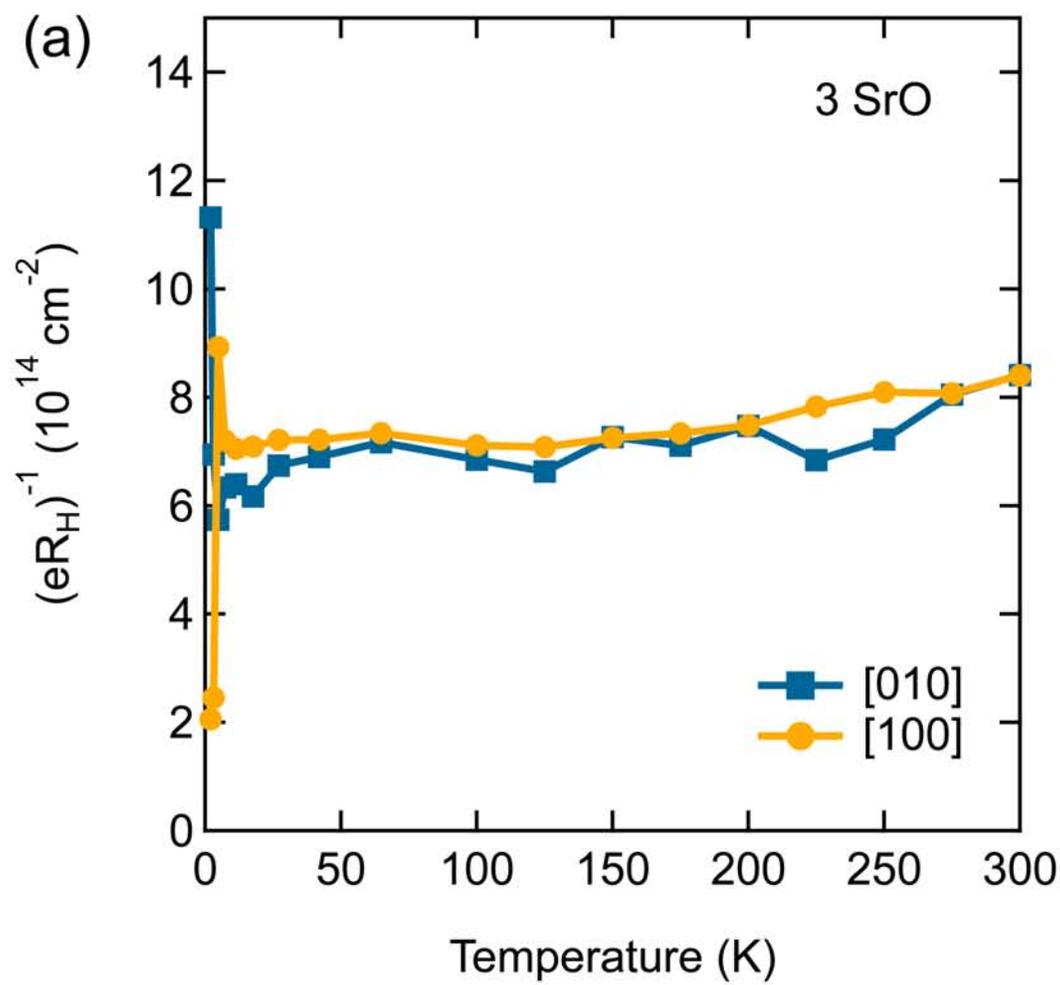
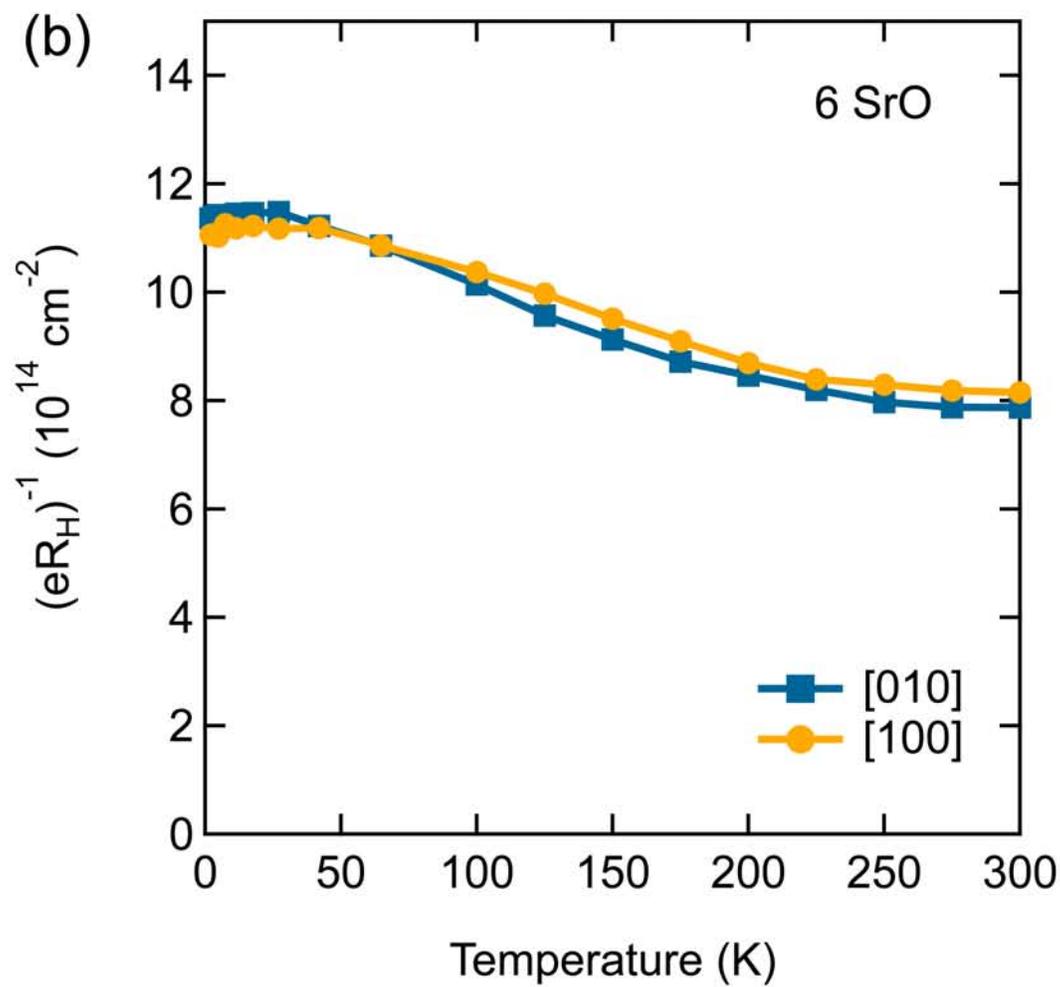

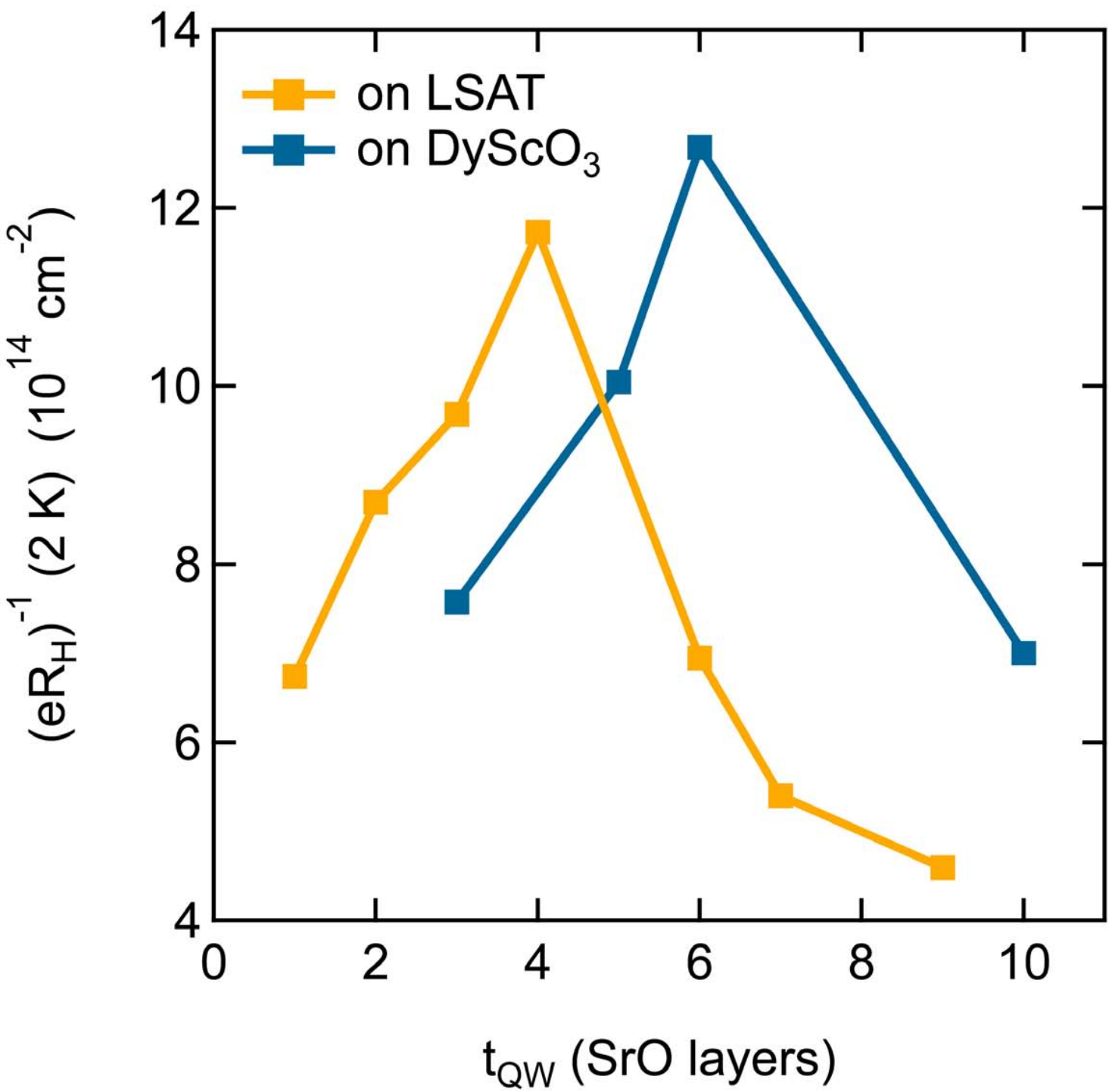